\documentclass[11pt]{article}

\newcommand{\nc}{\newcommand}
\nc{\eqr}[1]{(\ref{#1})}
\nc{\sref}[1]{\S~\ref{#1}}
\nc{\tref}[1]{Table~\ref{#1}}
\nc{\fref}[1]{Figure~\ref{#1}}
\nc{\cref}[1]{Chapter~\ref{#1}}
\nc{\bcenter}{\begin{center}}
\nc{\ecenter}{\end{center}}

\nc{\Htwo}{H^2(X,\real)}
\nc{\Htwoz}{H^2(X,\Z)}
\nc{\HttpiZ}{H^2(X,\tpi \Z )}
\nc{\Hp}{H^{2}_{\tp}}
\nc{\Hm}{H^{2}_{\tm}}
\nc{\imt}{\mbox{Im }\tau}
\nc{\imtpar}{(\mbox{Im }\tau )}
\nc{\onehalf}{\frac{1}{2}}
\nc{\onefour}{\frac{1}{4}}
\nc{\lra}{\longrightarrow}
\nc{\ra}{\rightarrow}
\nc{\mod}{\mbox{mod }}
\nc{\pf}{{\sc Proof: }}
\nc{\qbar}{\overline{q}}
\nc{\tbar}{\overline{\tau}}
\nc{\slz}{SL(2,\Z)}
\nc{\sltau}{\frac{a\tau +b}{c\tau +d}}
\nc{\tpi}{2\pi}
\nc{\tp}{\mbox{\tiny $+$}}
\nc{\tm}{\mbox{\tiny $-$}}
\nc{\vol}{\mbox{Vol }}
\nc{\ztwo}{\Z_{\mbox{\tiny $2$}}}
\nc{\zth}{\Z_{\mbox{\tiny $3$}}}
\nc{\dm}{\partial_{\mu}}
\nc{\dn}{\partial_{\nu}}
\nc{\dum}{\partial^{\mu}}
\nc{\dun}{\partial^{\nu}}
\nc{\Dm}{D_{\mu}}
\nc{\Dn}{D_{\nu}}
\nc{\Dum}{D^{\mu}}
\nc{\sqrg}{\sqrt{G}}
\nc{\Div}{\nabla\cdot}
\nc{\Curl}{\nabla\times}
\nc{\del}{\partial}
\nc{\tr}{\mbox{tr}}

\nc{\setall}{\setcounter{equation}{0}}
\nc{\setequation}{\setcounter{equation}{0}}

\def\sla#1{\raise.15ex\hbox{/}\kern-.57em #1}
\def\slas#1{\raise.15ex\hbox{/}\kern-.62em #1}
\nc{\tbyt}[4]{\left( \begin{array}{rr}
        #1 & #2 \\
        #3 & #4
        \end{array}\right)}
\nc{\abcd}{\left( \begin{array}{cc}
        a & b \\
        c & d
        \end{array}\right)}
\nc{\inner}[2]{\langle #1 , #2 \rangle}
\nc{\e}[1]{{\mbox e}^{#1}}
\nc{\met}[2]{g_{#1 #2}}
\nc{\oover}[1]{\frac{1}{#1}}
\nc{\wed}[2]{ #1 \wedge #2}
\nc{\bhat}[1]{\hat{\mbox{\boldmath $#1$}}}
\nc{\mbold}[1]{\mbox{\boldmath $#1$}}

\def\sCC{{\kern 0.27em\vrule height1.45ex width0.03em depth0em
          \kern-0.30em\rm C}}
\def\C{{\mathchoice
  {\sCC}
  {\sCC}
  {\kern 0.225em \vrule height1.05ex width0.025em depth0em \kern-0.25em \rm C}
  {\kern 0.180em \vrule height0.78ex width0.02em depth0em \kern-0.2em \rm C}
        }}
\def\sHH{{\rm I\kern-.16em{}H}}
\def\H{{\mathchoice
  {\sHH}
  {\sHH}
  {\rm I\kern-.13em{}H}
  {\rm I\kern-.13em{}H} }}
\def\sNN{{\rm I\kern-.16em{}N}}
\def\N{{\mathchoice
  {\sNN}
  {\sNN}
  {\rm I\kern-.12em{}N}
  {\rm I\kern-.10em{}N} }}
\def\sPP{{\rm I\kern-.16em{}P}}
\def\P{{\mathchoice
  {\sPP}
  {\sPP}
  {\rm I\kern-.12em{}P}
  {\rm I\kern-.10em{}P} }}
\def\sQQ{{\kern 0.27em \vrule height1.45ex width0.03em depth0em
          \kern-0.30em \rm Q}}
\def\Q{{\mathchoice
        {\sQQ}
        {\sQQ}
  {\kern 0.225em \vrule height1.05ex width0.025em depth0em \kern-0.25em \rm Q}
  {\kern 0.180em \vrule height0.78ex width0.020em depth0em \kern-0.20em \rm Q}
        }}
\def\sRR{{\rm I\kern-0.16em{}R}}
\def\R{{\mathchoice
  {\sRR}
  {\sRR}
  {\rm I\kern-0.12em{}R}
  {\rm I\kern-0.10em{}R} }}
\def\sZZ{{\rm Z\kern-0.32em{}Z}}
\def\Z{{\mathchoice
  {\sZZ}
  {\sZZ} 
  {\rm Z\kern-0.3em{}Z}     %.3
  {\rm Z\kern-0.25em{}Z} }}  %.25
\def\ZZZ{{\rm Z\kern-0.24em{}Z}}

\let\useblackboard=\iftrue
\newfam\black
\useblackboard
\font\blackboard=msbm10 scaled \magstep1
\font\blackboards=msbm7
\font\blackboardss=msbm5
\textfont\black=\blackboard
\scriptfont\black=\blackboards
\scriptscriptfont\black=\blackboardss
\def\Bbb#1{{\fam\black\relax#1}}
\else
\def\Bbb{\bf}
\fi

\def\yboxit#1#2{\vbox{\hrule height #1 \hbox{\vrule width #1
\vbox{#2}\vrule width #1 }\hrule height #1 }}
\def\fillbox#1{\hbox to #1{\vbox to #1{\vfil}\hfil}}
\def\ybox{{\lower 1.3pt \yboxit{0.4pt}{\fillbox{8pt}}\hskip-0.2pt}}

\def\QP{\Bbb{P}}
\def\QQ{\Bbb{Q}}

\nc{\vE}{\vec{E}}
\nc{\vB}{\vec{B}}

\nc{\cA}{{\cal A}}
\nc{\cF}{{\cal F}}
\nc{\cL}{{\cal L}}
\nc{\cN}{{\cal N}}
\nc{\cH}{{\cal H}}
\nc{\cO}{{\cal O}}
\nc{\cR}{{\cal R}}
\nc{\cg}{{\cal G}}

\nc{\bA}{{\bf A}}
\nc{\bB}{{\bf B}}
\nc{\bE}{{\bf E}}
\nc{\bI}{{\bf I}}
\nc{\bJ}{{\bf J}}
\nc{\bK}{{\bf K}}
\nc{\bR}{{\bf R}}
\nc{\bZ}{{\bf Z}}

\nc{\al}{\alpha}
\nc{\be}{\beta}
\nc{\ga}{\gamma}
\nc{\de}{\delta}
\nc{\ep}{\epsilon}
\nc{\n}{\nu}
\nc{\m}{\mu}
\nc{\sskip}{\vspace{5mm}}
\nc{\nskip}{\vspace{-2mm}}
\nc{\vs}[1]{\vspace{#1}}
\nc{\hs}{\hspace{1cm}}
\nc{\hshalf}{\hspace{5mm}}

\def\tone{t^{1}}
\def\tzero{t^{0}}
\def\uplus{u_{+}}
\def\uminus{u_{-}}

\def\matrix#1#2{\left( \begin{array}{#1} #2 \end{array} \right)}
\def\ga{\alpha}

\def\dd#1#2{{\partial #1 \over \partial #2}}

\newcommand{\pone}[1]{\left( #1 \right)}
\newcommand{\ptwo}[1]{\left[ #1 \right]}
\newcommand{\pthree}[1]{\left\{ #1 \right\}}

\ifx\undefined\square
  \def\square{\vrule width.6em height.5em depth.1em\relax}\fi
\def\qed{\ifhmode\unskip\nobreak\fi\quad
  \ifmmode\square\else$\m@th\square$\fi}

\newtheorem{theorem}{\bf Theorem}[section]

\newtheorem{definition}[theorem]{\bf Definition}

\newtheorem{lemma}[theorem]{\bf Lemma}

\newtheorem{conjecture}[theorem]{\bf Conjecture}

\renewcommand{\thefootnote}{\fnsymbol{footnote}}

\begin{document}

%{\flushright{\small MIT-CTP-3104 \\SLAC-PUB-8798\\SU-ITP-01/11\\
%}}

\begin{center}
{\LARGE\bf On a Conjecture of Givental}
\end{center}

\vspace{3mm}
\begin{center}
{\sc Jun S. Song\footnote{
E-mail:   jssong@alum.mit.edu.  }}\\ 
{\it Center for Theoretical Physics\\  Massachusetts 
Institute of Technology\\ Cambridge, MA 02139, U.S.A.}
\\
\vspace{4mm}
and
\vspace{4mm}
\\
{\sc Yun S. Song\footnote{E-mail: yss@stanfordalumni.org.
}
}\\ {\it Department of Physics \& SLAC \\ Stanford University \\
Stanford, CA 94305, U.S.A.}
\end{center}

\vspace{1cm}

\begin{abstract}
These brief notes record our puzzles and findings surrounding
Givental's recent conjecture which expresses higher genus Gromov-Witten
invariants in terms of the genus-0 data.  We limit our considerations
to the case of a complex projective line, whose Gromov-Witten invariants are
well-known and easy to compute.  We make some simple checks supporting
his conjecture.
\end{abstract}

\renewcommand{\thefootnote}{\arabic{footnote}}
\setcounter{footnote}{0}
\newpage
%%%%%%%%%%%%%%%%%%%%%%%%%%%%%%%%%%%%%%%%%%%%%%%%%
\section{Brief Summary}

These  notes
are brief sketches of our troubles and findings
surrounding a work of Givental \cite{Givental}.  

Let ${\cal F}_g$ be the generating function in the small phase space
for genus-$g$ Gromov-Witten
(GW) invariants of a manifold $X$ with a semi-simple Frobenius
structure on $H^*(X,\QQ)$.
Then, Givental's conjecture, whose equivariant counter-part he has
proved \cite{Givental}, is
        \begin{equation}
        e^{\sum_{g\geq 2} \lambda^{g-1}{\cal F}_g (t)} =\left.\left[ e^{
        {\lambda\over 2} \sum_{k,l\geq 0} \sum_{i,j} V^{ij}_{kl} \sqrt{\Delta_i} 
        \sqrt{\Delta_j}\partial_{q^i_k}\partial_{q^j_l}}\prod_j \tau( \lambda
        \Delta_j; \{q^j_n\}) \right]\right|_{q^j_n = T^j_n}\, ,\label{eq:master}
        \end{equation}
where $i,j = 1, \ldots \dim  H^*(X,\QQ)$; $\tau$ is the KdV tau-function
governing the intersection theory on the Deligne-Mumford space 
$\overline{{\cal M}}_{g,n}$;
and $V^{ij}_{kl}, \Delta_j,$  and $T^j_n$ are functions of the small phase
space coordinates $t \in H^*(X,\QQ)$ and are defined by solutions to the
flat-section equations associated with the genus-0 Frobenius structure
of $H^*(X,\QQ)$ \cite{Givental}.  This remarkable conjecture organizes
the higher 
genus GW-invariants in terms of the genus-0 data and the $\tau$-function
for a point.  The motivation for our work lies in verifying the conjecture
for $X= \QP^1$, which is the simplest example with a semi-simple Frobenius
structure on its cohomology ring and whose GW-invariants can be
easily computed. 

We have obtained two particular solutions to the flat-section equations
\eqr{eq:flat section}, an analytic one encoding the two-point descendant 
GW-invariants
of $\QP^1$ and a recursive one corresponding to Givental's fundamental 
solution.  According to Givental, both of
these two solutions are supposed to
yield the same data $V^{ij}_{kl}, \Delta_j,$  and $T^j_n$.  Unfortunately,
we were not able to produce the desired information using our analytic
solutions, but the recursive solutions do lead to sensible quantities which
we need.  Combined with an
expansion scheme which allows us to verify the conjecture at each order
in $\lambda$, we thus use our recursive solutions to check the conjecture
\eqr{eq:master} for $\QP^1$ up to order $\lambda^2$.  Already at this order,
we need to expand the differential operators in 
\eqr{eq:master} up to $\lambda^6$ and need to consider up to genus-3 free
energy in the $\tau$-functions, and the computations quickly become
cumbersome with increasing order.  We have managed to re-express the
conjecture 
for this case into a form which resembles the Hirota-bilinear relations, but
at this point, we have no insights into a general proof.   It is nevertheless
curious how the numbers work out, and we hope that our results 
would provide a humble support for Givental's master equation.

Many confusions still remain -- for instance, the discrepancy between
our analytic 
and recursive solutions.
As mentioned above, Givental's conjecture for $\QP^1$ can
be re-written in a form which resembles the Hirota-bilinear relations
for the KdV hierarchies (see \eqr{eq:hirota}). 
 It would thus be  interesting to speculate a
possible relation between
his conjecture and the conjectural Toda hierarchy for
$\QP^1$.  

We have organized our notes as follows:
in \S\ref{sec:review}, we review the canonical coordinates for $\QP^1$, to be
followed by our solutions to the flat-section equations in \S\ref{sec:sol}.
We conclude by presenting our checks in \S\ref{sec:checks}.
%\newpage

%%%%%%%%%%%%%%%%%%%%%%%%%%%%%%%%%%%%%%%%%%%%%%%%%
%\setall
\section{Canonical Coordinates for $\QP^1$.}\label{sec:review}
We here review the canonical coordinates $\{u_\pm\}$ for $\QP^1$
\cite{Dubrovin,DZ,ellipticGW}. 
Recall that a Frobenius structure on $H^*(\QP^1,\QQ)$ carries
 a flat pseudo-Riemannian metric $\langle\cdot ,\cdot \rangle$
defined by the Poincar\'e intersection pairing.
The canonical coordinates are defined by the property that they form
the basis of idempotents of the quantum cup-product,  denoted in the
present note by $\circ$.  The flat metric $\langle\cdot ,\cdot
\rangle$ is diagonal in the canonical coordinates, and following
Givental's notation, we define $\Delta_\pm :=1/ \langle
\partial_{u_\pm}, \partial_{u_\pm} \rangle$.

Let $\{t^\alpha\}\,, \alpha\in\{0,1\}$ be the flat coordinates of the
metric and let 
$\partial_\alpha := \partial/ \partial t^\alpha$.   
The quantum cohomology of $\QP^1$ is
\[
        \partial_0 \circ \partial_\alpha = \partial_\alpha
        \hspace{1cm} \mbox{and} \hspace{1cm}
        \partial_1 \circ \partial_1 = e^{t^1} \, \partial_0.
\]
The eigenvalues and eigenvectors of $\partial_1 \circ$ are
\[
        \pm \, e^{t^1/2} \hspace{1cm} \mbox{and} \hspace{1cm}
        ( \pm e^{t^1/4}\,\partial_0 + e^{-t^1/4} \,\partial_1),
\]
respectively.  So, we have
\[
        ( \pm e^{t^1/4}\,\partial_0 + e^{-t^1/4} \,\partial_1) \circ
        ( \pm e^{t^1/4}\,\partial_0 + e^{-t^1/4} \,\partial_1) =
        {\pm 2\,e^{t^1/4} }\,
         ( \pm e^{t^1/4}\,\partial_0 + e^{-t^1/4} \,\partial_1),
\]
which implies that
\[
        \frac{\partial}{\partial u_{\pm}} = \frac{
         \,\partial_0 \pm e^{-t^1/2} \,\partial_1}{2 }\, ,
\]
such that
\[
        \partial_{u_{\pm}} \circ \partial_{u_{\pm}} = \partial_{u_{\pm}}
        \hspace{1cm} \mbox{and} \hspace{1cm} 
        \partial_{u_{\pm}} \circ \partial_{u_{\mp}} = 0\, .
\]
We can solve for $u_{\pm}$ up to constants as 
        \begin{equation}
        u_{\pm} = t^0 \pm 2 \, e^{t^1/2}\, . \label{eq:canonical u}
        \end{equation}

To compute $\Delta_{\pm}$, note that
\[
{1\over \Delta_{\pm}} :=  \langle \partial_{u_\pm}, \partial_{u_\pm} \rangle =
        \pm {1\over 2 e^{t^1/2}}\, .
\]
The two bases are related by
\[
        \partial_0 = \partial_{u_{+}} + \partial_{u_{-}}
        \hspace{1cm} \mbox{and} \hspace{1cm}
        \partial_1 =  \,e^{t^1/2} \, ( \partial_{u_{+}} 
        - \partial_{u_{-}})\, .
\]
Define an orthonormal basis by $f_{i}=\Delta_i^{1/2}\dd{\ }{u_{i}}$.
Then the transition matrix $\Psi$ from  \{$\dd{\ }{t_\ga}$\} to \{$f_i$\}
is given by
        \begin{equation}
        \Psi^{\ i}_\ga = {1\over \sqrt{2}} \matrix{cc}{e^{-\tone/4} &
        -i\, e^{-\tone/4}\\e^{\tone/4} & i\, e^{\tone/4}}
        = \matrix{cc}{\Delta_+^{-1/2}&\Delta_{-}^{-1/2}\\{1\over 2}\Delta_+^{1/2}
        &{1\over 2}\Delta_{-}^{1/2} } \, ,
        \label{eq:transitionM}
        \end{equation}
such that
\[
\dd{\ }{t_\ga} = \sum_i \Psi^{\ i}_{\ga}\ f_i.
\]
We will also need the inverse of \eqr{eq:transitionM}:
\begin{equation}
(\Psi^{-1})_{i}^{\ \ga} = {1\over \sqrt{2}} \matrix{cc}{e^{\tone/4} &
 e^{-\tone/4}\\i\, e^{\tone/4} & -i\, e^{-\tone/4}} =
\matrix{cc}{{1\over 2}\Delta_+^{1/2}&\Delta_+^{-1/2}\\
 {1\over 2}\Delta_{-}^{1/2}&\Delta_{-}^{-1/2}} .
\label{eq:psiInverse}
\end{equation}

%%%%%%%%%%%%%%%%%%%%%%%%%%%%%%%%%%%%%%%%%%%%%%%%
%\setall
\section{Solutions to the Flat-Section Equations} \label{sec:sol}
The relevant data $V^{ij}_{kl}, \Delta_j$ and $T^j_n$ are extracted
from the solutions to the flat-section equations of the genus-0
Frobenius structure for $H^*(\QP^1,\QQ)$.  We here find two particular
solutions.  The analytic solution correctly encodes the two-point
descendant GW-invariants, while the recursive solution is used in
the next section to verify Givental's conjecture.

\subsection{Analytic Solution}
\label{subsection:analytic}
The genus-0 free energy for $\QP^1$ is
        \[
        {\cal F}_0 = {1\over2}{(t^0)^2 t^1} + e^{t^1}.
        \]
Flat sections $S_\alpha$ of $T H^*(\QP^1,\QQ)$ satisfy the equations
        \begin{equation}
        z\, \partial_{\alpha}  \, S_{\beta} =
        {\cal F}_{\alpha\beta\mu} \, g^{\mu\nu}
        \,S_{\nu}\ , \label{eq:flat section}
        \end{equation}
where $z\neq 0$ is an arbitrary parameter and ${\cal
        F}_{\alpha\beta\mu} := \partial^3 {\cal F}/ \partial t^\alpha
\partial t^\beta \partial t^\mu$.
The only non-vanishing components of ${\cal F}_{\alpha\beta\mu}$ are 
\[
        \cF_{001} =1  \hspace{1cm} \mbox{and}
        \hspace{1cm}  \cF_{111} = e^{t^1}\ . 
\]
Hence, we find that the  general solutions to the flat-section
equations \eqr{eq:flat section} are
        \begin{equation}
        S_0 =  e^{t^0/z}\, 
        \left[ c_1 \, I_0(2\, e^{t^1/2}/z)  - c_2 \, K_0 (2 \,
        e^{t^1/2}/z ) \right] \label{eq:szero}
        \end{equation}
and
\[
        S_1 = e^{t^0/z}\,  e^{t^1/2} \left[ c_1 \, I_1( 2 e^{t^1/2}
        /z) + c_2 \, 
          K_1(2 e^{t^1/2}/z) \right] \, ,
\]
where $I_n(x)$ and $K_n(x)$ are modified Bessel functions, and $c_i$
are integration constants which may depend on $z$.

We would now like to find two particular solutions
corresponding to the following Givental's expression:
        \begin{equation}
        S_{\alpha\beta}(z) = g_{\alpha\beta} + \sum_{n\geq 0, (n,d)\neq
        (0,0)} {1\over n!}\, \langle \phi_{\alpha} \cdot { \phi_{\beta}\over
        z-\psi} \cdot( t^0 \phi_0 + t^1 \phi_1)^n \rangle_d\, , \label{eq:sab}
        \end{equation}
where
$S_{\alpha\beta}$ denotes the $\alpha$-th component of the $\beta$-th
solution.
Here, $\{\phi_\alpha\}$ is a homogeneous basis of $H^*(\QP^1,\QQ)$,
$g_{\alpha\beta}$ is the intersection paring $\int_{\QP^1} \phi_\alpha
\cup \phi_\beta$  and $\psi\in H^2(\overline{M}_{0,n+2}(\QP^1,d),\QQ)$
is the first Chern class of the universal cotangent line bundle over
the moduli space\break $\overline{M}_{0,n+2}(\QP^1,d)$.  
In order to find the particular solutions, we compare our general
solution \eqr{eq:szero} with 
the 0-th components of $S_{0\beta}$ in \eqr{eq:sab} 
{\it at the origin of the phase space}.
The two-point functions appearing in \eqr{eq:sab} have been computed
at the origin in \cite{song} and have the following forms:
        \begin{equation}
        S_{00}|_{t^\alpha=0} = - \sum_{m=1}^{\infty}{1\over z^{2m+1}}
        {2\, d_m \over (m!)^2}\ ,\ \  \mbox{where } d_m =
        \sum^m_{k=1} 1/k\, , \label{eq:s00} 
        \end{equation}
and
        \begin{equation}
        S_{01}|_{t^\alpha=0} = 1 + \sum_{m=1}^{\infty} {1\over z^{2m}} 
        {1\over (m!)^2}. \label{eq:s01}
        \end{equation}
Using the standard expansion of the modified Bessel function $K_0$, 
we can evaluate \eqr{eq:szero} at the origin of the phase space to
be
        \begin{eqnarray}
        c_1 \, I_0\left({2\over z}\right)  - c_2 \, K_0 \left({2\over
        z}  \right) 
        &=& c_1\, I_0\left({2\over z}\right) -c_2 \left[- \left(-\log (z) + 
        \gamma_{\mbox{\tiny $E$}}\right) I_0\left({2\over z}\right)
          + \sum_{m=1} {c_m\over
        z^{2m} (m!)^2}\right]\, ,\nonumber \\ \label{eq:small}
        \end{eqnarray}
where $\gamma_{\mbox{\tiny $E$}}$ is Euler's constant.
Now matching \eqr{eq:small} with \eqr{eq:s00} gives
\[
        c_1  = - c_2 \log (1/z) - c_2 \gamma_{\mbox{\tiny $E$}}
        \hspace{1cm} \mbox{and} 
        \hspace{1cm} c_2 = {2\over z}\, ,
\]
while noticing that \eqr{eq:s01} is precisely the expansion of 
$I_0(2/z)$ and demanding that our general solution coincides with
\eqr{eq:s01} at the origin yields
\[
        c_1 =1 \hspace{1cm} \mbox{and}
        \hspace{1cm} c_2 =0 \ . 
\]
To recapitulate, we have found
\begin{eqnarray*}
        S_{00} &=& -{2 e^{t^0/z}\over z} \, 
        \left[ \left(\gamma_{\mbox{\tiny $E$}}-\log(z)\right) \,
        I_0\left({2\, e^{t^1/2}\over  
        z}\right)  + \, K_0 \left({2 \,
        e^{t^1/2}\over z }\right) \right]\, ,\\
    S_{10} &=& {2 e^{t^0/z}\,  e^{t^1/2} \over z}
        \left[K_1\left({2 e^{t^1/2}\over 
        z}\right) - \left( \gamma_{\mbox{\tiny $E$}}-\log ({z})\right)
        I_1\left( {2 e^{t^1/2} 
        \over z}\right) \right]\, ,\\
        S_{01} &=&  e^{t^0/z} \, I_0\left({2\, e^{t^1/2}\over
        z}\right)\, , \\
        S_{11} &=&  e^{t^0/z}\,  e^{t^1/2}\, I_1\left( {2 e^{t^1/2}
        \over z}\right)\, .
\end{eqnarray*}
We have checked that these solutions correctly reproduce the corresponding
descendant Gromov-Witten invariants obtained in \cite{song}. 

If the inverse transition matrix in \eqr{eq:psiInverse} is used to relate
the matrix elements $S^{\ i}_{\alpha}$ to $S_{\alpha\beta}$ 
as $S_\alpha^{\ i} = S_{\alpha\beta} ((\psi^{-1})^t)^\beta_{\ j}\,
\delta^{ji}$, then we should have 
        \begin{equation}
        S^{\ \pm}_{\alpha} = \sqrt{\pm 2} \, e^{t^1/4}\, \left(
        {1\over 2} S_{\alpha 0} \pm {e^{-t^1/2}\over 2} S_{\alpha 1}\right).
      \label{eq:S-alpha-pm}
        \end{equation}

%%%%%%%%%%%%%%%%%%%%%%%%%%%%%%%%%%%%%%%%%%%%%%%%

\subsection{Recursive Solution}
\label{subsection:recursion}
In \cite{ellipticGW,Givental}, Givental has shown that 
near a semi-simple point, the flat-section equations
\eqr{eq:flat section} have a fundamental solution given by 
\[
S_\ga^{\ i} = \Psi_\ga^{\ j} (R_0 + z R_1 + z^2 R_2 + \cdots + z^n R_n +
\cdots)_{jk}
\ptwo{\exp(U/z)}^{ki}\, ,
%\equiv \Psi {\bf R}\exp(U/z),
\]
where $R_n=(R_n)_{jk}$, $R_0=\delta_{jk}$ and $U$ is the diagonal
matrix of canonical coordinates.
The matrix $R_1$ satisfies the relations 
        \begin{equation}
        \Psi^{-1}\dd{\Psi}{\tone} = [\dd{U}{\tone}, R_1] \label{eq:Psi-1Psi}
        \end{equation}
and
        \begin{equation}
        \left[\dd{R_1}{\tone} + \Psi^{-1}\left(\dd{\Psi}{\tone}\right)
        R_1\right]_{\pm\pm} = 0 \, ,\label{eq:R'} 
        \end{equation}
which we use to find its expression.  From the transition matrix given in
\eqr{eq:transitionM} we see that
\[
\Psi^{-1}\dd{\Psi}{\tone} = {1\over 4}\matrix{cc}{0&i\\-i&0},
\]
while taking the $(+-)$ component of the relation \eqr{eq:Psi-1Psi} gives
\[
{i \over 4} = \dd{U_{++}}{\tone} (R_1)_{+-}
-(R_1)_{+-}\dd{U_{--}}{\tone} = 2 e^{\tone/2} (R_1)_{+-} \, ,
\]
where in the last step we have used the definition  
\eqr{eq:canonical u} of canonical coordinates.  
We therefore have
\[
  (R_1)_{+-} = {i \over 8} e^{-\tone/2}\, ,
\]
and similarly considering the $(-+)$ component of \eqr{eq:Psi-1Psi} gives
\[
(R_1)_{-+} = {i \over 8} e^{-\tone/2}.
\]
The diagonal components of $R_1$ can be obtained from \eqr{eq:R'},
which implies that
        \[
        \dd{(R_1)_{++}}{\tone} =
        (R_1)_{+-}\dd{U_{--}}{\tone}(R_1)_{-+}
        - \dd{U_{++}}{\tone} (R_1)_{+-}(R_1)_{-+} =
        {\exp(-\tone/2)\over 32}= - 
        \dd{(R_1)_{--} }{\tone}.
        \]
Hence,  $(R_1)_{++}= -\exp(-\tone/2)/16 = - (R_1)_{--}$ and the matrix
$R_1$ can be written as
        \begin{equation}
        (R_1)_{jk} =
        {1\over16}e^{-\tone/2}\matrix{cc}{-1&2i\\2i&1}.
        \label{eq:R_1}
        \end{equation}
In general, the matrices $R_n$ satisfy the recursion relations
\cite{ellipticGW}
\[
%\pone{\dd{\ }{\tone} + \Psi^{-1}\dd{\Psi}{\tone}} R_n=[\dd{U}{\tone}, R_{n+1}]
        \pone{d + \Psi^{-1} d\Psi} R_n = [d U, R_{n+1}]\ ,
\]
which, for our case, imply the following set of equations:
        \begin{eqnarray}
        \dd{R_n}{\tzero} &=& 0 \ , \label{eq:dR/dt0}\\
        \dd{(R_n)_{++}}{\tone} &=& - {i\over 4} (R_n)_{-+} \ , \label{eq:R++}\\
        (R_{n+1})_{-+} &=& -{1\over2}e^{-\tone/2}
        \ptwo{\dd{(R_n)_{-+}}{\tone}  - {i\over 4} (R_n)_{++}}\ ,
        \label{eq:R-+}\\
        \dd{(R_n)_{--}}{\tone} &=& {i\over 4} (R_n)_{+-}\ , \label{eq:R--}\\
        (R_{n+1})_{+-} &=& {1\over2}e^{-\tone/2}
        \ptwo{\dd{(R_n)_{+-}}{\tone}  + {i\over 4} (R_n)_{--}}\ .
        \label{eq:R+-}
        \end{eqnarray}

\begin{lemma}
For $n\ge 1$, the matrices $R_n$ in the fundamental solution
are given by
        \begin{equation}
        (R_n)_{ij} = {(-1)^n\over (2n-1)} {\ga_n\over 2^n}\, e^{-n\tone/2}
        \matrix{cc}{-1&(-1)^{n+1}\, 2n\, i \\ 2n\,i&(-1)^{n+1}},
        \label{eq:Rsolution} 
        \end{equation}
where
\[
\alpha_n = (-1)^n {1\over 8^n n!} \prod_{\ell=1}^{n} (2\ell-1)^2\ , \
\alpha_0=1.
\]
These solutions
satisfy the unitarity condition 
\[\mbox{\boldmath $R$}(z)\mbox{\boldmath $R$}^t(-z) :=
(1 + z R_1 + z^2 R_2 + \cdots + z^n R_n + \cdots)
(1 - z R_1^t + z^2 R_2^t + \cdots + (-1)^n z^n R_n^t + \cdots)
= 1
\]
and the homogeneity condition and, thus, are unique. \label{lemma:R}
\end{lemma}
{\sc Proof:}
For $n=1$, $\ga_1=-1/8$ and \eqr{eq:Rsolution} is  equal
to the correct solution \eqr{eq:R_1}.
The proof now follows by an induction on $n$.
Assume that \eqr{eq:Rsolution} holds true up to and including $n=m$.
Using the fact that
\[
\ga_{m+1} = - {(2m+1)^2\over 8(m+1)} \ga_{m},
\]
we can show that $R_{m+1}$ in \eqr{eq:Rsolution} satisfies the
relations \eqr{eq:R++}--\eqr{eq:R+-} as well as \eqr{eq:dR/dt0}.

To check unitarity, consider the $z^k$-term $P_k := \sum^{k}_{\ell=0}
(-1)^\ell\, R_{k-\ell}\,R^t_\ell$ in  
$\mbox{\boldmath $R$}(z)\mbox{\boldmath $R$}^t(-z) = \sum_{k=0} {P_k
z^k}$.  As shown by Givental,  the 
equations satisfied by the matrices $R_n$ imply that the off-diagonal
entries of $P_k$ vanish.  As a result, combined with the anti-symmetry
of $P_k$ for odd $k$, we see that $P_k$ vanishes for $k$ odd.  Hence, we
only need to show that for our solution, $P_{k}$ vanishes for all
positive even $k$ as well.   To this end, we note that Givental has
also deduced from the 
equation $dP_k + [\Psi^{-1} d \Psi,P_k] = [dU, P_{k+1}]$ that the
diagonal entries of 
$P_k$ are constant.  The expansion of $P_{2k}$ is 
\[
        P_{2k} = R_{2k} + R^t_{2k} + \cdots\, , \label{eq:P2k} 
\]
where the remaining terms are products of $R_{\ell}$, for $\ell < 2k$.
Now, we proceed inductively.  We first note that $R_1$ and $R_2$ given in
\eqr{eq:Rsolution} satisfy the condition $P_2 =0$, and assume that
$R_{\ell}$'s  in \eqr{eq:Rsolution} for $\ell < 2k$ satisfy $P_{\ell}
=0$.  Then, since the off-diagonal entries of $P_n$ vanish for all $n$,
the expansion of $P_{2k}$ is of the form
\[
        P_{2k} = A \, e^{-2k\,t^1/2} + B ,
\]
where $A$ is a constant diagonal matrix resulting 
from substituting our solution
\eqr{eq:Rsolution} and $B$ is a possible diagonal matrix of
integration constants for $R_{2k}$.  But, since the diagonal entries
of $P_{n}$ are constant for all $n$, we know that $A=0$.  We finally choose the
integration constants to be zero so that $B=0$, yielding $P_{2k}=0$.
Hence, the matrices in our solution \eqr{eq:Rsolution} satisfy the unitarity
condition and are manifestly homogeneous.  It then follows by the
proposition in \cite{Givental} that our solutions $R_n$ are  unique.
\hfill$\qed$

\vspace{.5cm}
Let $\mbox{\boldmath $R$} :=  (R_0+z R_1+z^2 R_2+\cdots  + z^n R_n + \cdots)$.
Then, we can use the matrices $R_n$ from Lemma~\ref{lemma:R} to find 
\begin{eqnarray}
S_0^{\ +} 
&=& \pone{\mbox{\boldmath $R$}_{++} - i\, \mbox{\boldmath $R$}_{-+}} {\exp({\uplus/z})
\over \sqrt{\Delta_{+}}}\nonumber \\
&=&\ptwo{1 + \sum_{n=1}^{\infty} {\ga_n\over 2^n} \exp\pone{-n
\tone\over2} (-z)^n}
 {\exp({\uplus/z}) \over \sqrt{\Delta_{+}}}\ ,  \label{eq:S0+} \\
S_0^{\ -} &=& 
\pone{\mbox{\boldmath $R$}_{--} + i\, \mbox{\boldmath $R$}_{+-}}
{\exp({\uminus/z}) 
\over \sqrt{\Delta_{-}}}\nonumber \\
&=&\ptwo{1 + \sum_{n=1}^{\infty} (-1)^n {\ga_n\over 2^n} \exp\pone{-n
\tone\over2} (-z)^n}
 {\exp({\uminus/z}) \over \sqrt{\Delta_{-}}}\ ,  \label{eq:S0-}\\
%%%%%%%%%%%%%%%%
S_1^{\ +} &=&
\pone{\mbox{\boldmath $R$}_{++} + i\, \mbox{\boldmath $R$}_{-+}}
{\sqrt{\Delta_{+}}\over 2} \exp({\uplus/z})
\nonumber \\
&=&\ptwo{1 - \sum_{n=1}^{\infty} {(2n+1)\over (2n-1)}{\ga_n\over 2^n}
\exp\pone{-n
\tone\over2} (-z)^n} {\sqrt{\Delta_{+}}\over 2}
 \exp({\uplus/z}) \ ,  \label{eq:S1+}\\
%%%%%%%%%%%%%%%%
S_1^{\ -} &=&
\pone{\mbox{\boldmath $R$}_{--} - i\, \mbox{\boldmath $R$}_{+-}}
{\sqrt{\Delta_{-}}\over 2} \exp({\uminus/z}) \nonumber \\
&=&\ptwo{1 - \sum_{n=1}^{\infty} (-1)^n {(2n+1)\over (2n-1)}
{\ga_n\over 2^n} \exp\pone{-n
\tone\over2} (-z)^n} {\sqrt{\Delta_{-}}\over 2}
\exp({\uminus/z}) \ .  \label{eq:S1-}
\end{eqnarray}
%%%%%%%%%%%%%%%
Using the above expressions for $S_{\ga}^{\ i}(z)$, we
can also find  $V^{ij}(z,w)$, which is given by the expression
\[
V^{ij}(z,w) := {1\over z+w}\ [S_\mu^{\ i}(w)]^t\,
[g^{\mu\nu}]\,[S_\nu^{\ j}(z)] .
\]
\noindent
If we define
\[
A_{p,q} :=
{(4p\,q-1)\over(2p-1)(2q-1)} {\ga_p\,\ga_q\over 2^{p+q}}
e^{-(p+q)\tone\over2}
\]
and
\[
B_{p,q} := {2(p-q)\over(2p-1)(2q-1)}
{\ga_p\,\ga_q\over 2^{p+q}}e^{-(p+q)\tone\over2},
\]
then after some algebraic manipulations we obtain
\begin{eqnarray}
V^{++}(z,w) &=& {e^{u_{+}/w+u_{+}/z}}
\pthree{{1\over z+w}  +
\sum_{k,l=0}^{\infty}
\ptwo{ \sum_{n=0}^k (-1)^n
A_{l+n+1, k-n}}
(-1)^{k+l}\, w^k z^l }, \label{eq:V++} \\
%%%%%%%
V^{--}(z,w) &=& {e^{u_{-}/w+u_{-}/z}}
\pthree{{1\over z+w}  -
\sum_{k,l=0}^{\infty}
\ptwo{ (-1)^{k+l} \sum_{n=0}^k (-1)^n
A_{l+n+1, k-n}} (-1)^{k+l}\, w^k z^l }\, ,\nonumber \\
%%%%%%%
V^{+-}(z,w) &=& e^{u_{+}/w+u_{-}/z}
\pthree{\sum_{k,l=0}^{\infty}
\ptwo{i\, (-1)^l \sum_{n=0}^k
B_{l+n+1, k-n}}\
(-1)^{k+l}\,  w^k z^l }\, , \label{eq:V+-} \\
%%%%%%%
V^{-+}(z,w) &=& e^{u_{-}/w+u_{+}/z}
\pthree{\sum_{k,l=0}^{\infty}
\ptwo{i\, (-1)^k \sum_{n=0}^k
B_{l+n+1, k-n}}\
(-1)^{k+l} \, w^k z^l }\, .\nonumber 
\end{eqnarray}

%%%%%%%%%%%%%%
%
%%%%%%%%%%%%%%

\subsection{A Puzzle}

Incidentally, we note that in the asymptotic limit $z\rightarrow 0$,
\[
S_0^{\ +} = \Re\ptwo{\sqrt{2\pi \over z} e^{\tzero/z}
I_0\pone{2e^{\tone/2}\over z}}
\]
and
\[
S_0^{\ -} = -i \sqrt{2\over\pi z} e^{\tzero/z}
K_0\pone{2e^{\tone/2}\over z}
\]
reproduce the expansions in \eqr{eq:S0+} and \eqr{eq:S0-}. 
This is in contrast to what was expected from the discussion leading
to  \eqr{eq:S-alpha-pm}. 
Despaired of matching the two expressions, it seems to us that 
the analytic correlation functions obtained  in
\S\ref{subsection:analytic} do not encode the right information that
appear in Givental's conjecture.   In the following section, we will
use the recursive
solutions from \S\ref{subsection:recursion} to check Givental's conjectural
formula at low genera.

%%%%%%%%%%%%%%%%%%%%%%%%%%%%%%%%%%%%%%%%%%%%%%%%
%\setall
\section{Checks of the Conjecture at Low Genera}\label{sec:checks}
The $T^i_n$ that appear in Givental's formula \eqr{eq:master} are defined by the
equations \cite{Givental}
\[
S_0^{\ \pm} := \left[ 1- \sum_{n=0}^\infty T_n^\pm (-z)^{n-1} \right]
{\exp(u_\pm/z)\over \sqrt{\Delta_\pm}}\, .
\]
From the computations of $S^{\ +}_0$ and $S^{\ -}_0$ in \eqr{eq:S0+} and
\eqr{eq:S0-}, respectively, one can extract $T^i_n$ to be
%%%%%%%%%%%%%%%%%%%
\begin{eqnarray*}
T^{+}_n &=& 
 \left\{ 
    \begin{array}{cl} 
      0 \, , & n=0,1\, , \\ \displaystyle
      -\, {\alpha_{n-1}\over 2^{n-1}}\, \exp\left[ 
          {-(n-1) t^1 \over 2}\right]\, , \hspace*{1.35cm} & n\geq 2\, , 
    \end{array} 
 \right.\\
\nonumber\\
T^{-}_n &=&
 \left\{ 
    \begin{array}{cl} 
      0 \, , & n=0,1\, , \\ \displaystyle
        -\, (-1)^{n-1}  {\alpha_{n-1}\over 2^{n-1}} \, \exp
        \left[  { -(n-1) t^1 \over 2}\right]\, , & n\geq 2\, .
    \end{array}
 \right.
\end{eqnarray*}
%%%%%%%%%%%%%%%%%%%
Notice that 
\begin{equation}
        T^{-}_n = (-1)^{n-1}\, T^{+}_n\, .\label{eq:relations}
\end{equation}
The functions $V^{ij}_{kl}$ are defined\footnote{There seems to
be a misprint in the original formula for $V^{ij}_{kl}$ in 
\cite{Givental}, i.e. we believe that $w$ and $z$ should be exchanged, 
as in our expression here.} 
by the expansion \cite{Givental}
\[
V^{ij}(z,w) = e^{u^i/w + u^j/z} \left[
{\delta^{ij} \over z+w} + \sum_{k,l=0}^\infty
(-1)^{k+l} \, V^{ij}_{kl}\, w^k z^l \right]\, ,
\]
and from \eqr{eq:V++} and \eqr{eq:V+-} we see that
\begin{eqnarray*}
                V^{++}_{kl} &=& \sum^k_{n=0} (-1)^n A_{l+n+1, k-n} = 
                        \sum^k_{n=0} {(-1)^n (4 (l+n +1)(k-n) -1) \over
                      (2l + 2n +1) (2k -2 n -1)}\,
                      T^{+}_{l+n+2}T^{+}_{k-n+1}\, , \\
                V^{+-}_{kl} &=&  i(-1)^l  \sum^{k}_{n=0} B_{l+n+1,k-n}
                        =  i(-1)^l\,\sum^k_{n=0} { 2(l+2 n +1 - k) \over
                  (2l + 2n +1) (2k -2 n -1) }\, T^{+}_{l+n+2}T^{+}_{k-n+1} .
\end{eqnarray*}

Now, the $\tau$-function for the intersection theory on the
Deligne-Mumford moduli space $\overline{\cal M}_{g,n}$  of stable
curves is defined by
\[
        \tau (\lambda; \{q_k\}) = \exp \left( \sum^{\infty}_{g=0}\,
        \lambda^{g-1} \, {\cal F}^{\mbox{\tiny pt}}_g (\{q_k\})\right)
\]
and has the following nice scaling invariance: consider the scaling of
the phase-space variables $q_k$ given by
\begin{equation}
        q_k \mapsto s^{k-1}\, q_k \label{eq:scaling}
\end{equation}
for some constant $s$.  Then, since a non-vanishing intersection
number $\langle \tau_{k_1} \cdots \tau_{k_n}\rangle$ must satisfy
\[
        \sum^n_{i=1}( k_i - 1) = \dim  (\overline{\cal M}_{g,n}) - n =
        3g-3  \, ,
\]
we see that under the transformation \eqr{eq:scaling}, 
the genus-$g$ generating function ${\cal F}^{\mbox{\tiny pt}}_g$ must behave as
\[
        {\cal F}^{\mbox{\tiny pt}}_g ( \{s^{k-1} \, q_k\}) =
        (s^3)^{g-1} \, {\cal F}^{\mbox{\tiny pt}}_g  (\{q_k\}) .
\]
Hence, upon scaling the ``string coupling constant'' $\lambda$ to
$s^{-3}\, \lambda$, we see that
\begin{equation}
        \tau (s^{-3} \lambda;\, \{s^{k-1}q_k\}) = \tau (\lambda; \{q_k\})\,
        . \label{eq:tau-scale}
\end{equation}
Now, consider the function
\begin{equation}
        F(\{q^{+}_n\}, \{q^{-}_n\}) := 
         \left[ \e{{\lambda \over 2} \sum_{k,l\geq 0}\sum_{i,j\in \{\pm \}}
                 V^{ij}_{kl} \sqrt{\Delta_i}\sqrt{
\Delta_j} \partial_{q^i_k} \partial_{q^j_l}}\, \tau(\lambda\Delta_{+};
\{q^{+}_n\}) \tau(\lambda\Delta_{-}; \{q^{-}_n\})\right]
%\vrule_{q^{+}_n=\beta , \, q^{-}_k=\beta^{-}_k} 
\, .\label{eq:Givental}
\end{equation}
Then, since the Gromov-Witten potentials of $\QP^1$ for $g\geq 2$ all vanish, 
Givental's conjectural formula for $\QP^1$ is
        \[
        F(\{T_n^{+}\}, \{T_n^{-}\}) = 1 \, ,
        \]
where it is understood that one sets $q_k^i=T_k^i$ after taking the
        derivatives with respect to $q_k^i$.
Since $T^{+}_n$ and  $T^{-}_n$ are related by \eqr{eq:relations},
let us rescale $q^{-}_{k} \mapsto (-1)^{k-1} q^{-}_{k}$ in
\eqr{eq:Givental}.  Then, since $\Delta_{+} = - \Delta_{-}$, we
observe from \eqr{eq:tau-scale} that 
{\small       \begin{eqnarray*}
 F(\{T_n^{+}\}, \{T_n^{-}\})&=&  \left\{ \exp\left[ {\lambda \over 2}
 \Delta_{+}\sum_{k,l\geq 0}\left(  V^{++}_{kl}
 \partial_{q^{+}_k} \partial_{q^{+}_l}  +  i (-1)^{l-1}\,V^{+-}_{kl} 
 \partial_{q^{+}_k} \partial_{q^{-}_l}+ i (-1)^{k-1}\,V^{-+}_{kl} 
 \partial_{q^{-}_k} \partial_{q^{+}_l}\right.\right.\right.\nonumber\\ 
  &&\left.\left.\left. \left. -\,(-1)^{k+l} V^{--}_{kl}
 \partial_{q^{-}_k} \partial_{q^{-}_l}  \right) \rule{0mm}{0.68cm}\right] \,
\tau(\lambda\Delta_{+}; 
\{q^{+}_n\}) \tau(\lambda\Delta_{+}; \{q^{-}_n\})\right\}
        \right|_{q^{+}_n\, , q^{-}_n = 
T^{+}_n}. 
        \end{eqnarray*}}
But, the $V^{ij}_{kl}$ satisfy the relations $V^{--}_{kl} = -
(-1)^{k+l}V^{++}_{kl}$ and $V^{+-}_{kl} = V^{-+}_{lk}$, so 
       \begin{eqnarray}
 F(\{T_n^{+}\}, \{T_n^{-}\})
 &=&  \left\{ \exp\left[ {\lambda \over 2} \Delta_{+} \sum_{k,l\geq 0}
 \left( \, V^{++}_{kl}
( \partial_{q^{+}_k} \partial_{q^{+}_l}+ \partial_{q^{-}_k}
\partial_{q^{-}_l} ) \ + \right.\right.\right.\nonumber\\ 
  &&\left.\left.\left. \left.  +\, 2i\, (-1)^{l-1}\,V^{+-}_{kl} 
 \partial_{q^{+}_k} \partial_{q^{-}_l}  \right) \rule{0mm}{0.69cm}\right] \,
\tau(\lambda\Delta_{+}; 
\{q^{+}_n\}) \tau(\lambda\Delta_{+}; \{q^{-}_n\})\right\}  
\right|_{q^{+}_n\, , q^{-}_n =T^{+}_n}.\nonumber \\ \label{eq:conj-simple}
        \end{eqnarray}
Now, consider the following transformations of the variables:
        \[
        q^{+}_k = x_k + y_k \hspace{1cm} \mbox{and} \hspace{1cm} 
        q^{-}_k = x_k - y_k  
        \]
so that
        \[
        \partial_{q^{+}_k} = \onehalf \left(\partial_{x_k} +
\partial_{y_k} \right)\hspace{1cm} \mbox{and} \hspace{1cm}  
        \partial_{q^{-}_k} = \onehalf \left(\partial_{x_k} -
\partial_{y_k} \right).
        \]
Then, in these new coordinates, \eqr{eq:conj-simple} becomes 
\[
                F(\{T_n^{+}\},\{ T_n^{-}\}) = G(\{T^{+}_n \}, \{0\}),
\]
where the new function $G(\{x_k\},\{y_k\})$ is defined\footnote{We
have simplified the expression by noting that the mixed derivative terms cancel
because of the identity $V^{+-}_{kl} = (-1)^{k-l}\, V^{+-}_{lk}.$} by
        {\small \begin{eqnarray}
G(\{x_n\},\{y_n\}) &=&   \exp\left[ {\lambda \over 4}\Delta_{+}\sum_{k,l\geq 0}
\left( V_{kl} \partial_{x_k}\partial_{x_l}+ 
 W_{kl} \partial_{y_k}\partial_{y_l}\right) \right] 
\tau(\lambda\Delta_{+}; \{x_n+y_n\}) \tau(\lambda\Delta_{+};
\{x_n-y_n\}) ,\nonumber\\\label{eq:hirota}
         \end{eqnarray}}
where
\begin{eqnarray*}
                V_{kl}\, &:=& \, V^{++}_{kl} + i (-1)^{l-1}\, V^{+-}_{kl}\ , \\
                W_{kl}\, &:=&\, V^{++}_{kl} - i (-1)^{l-1}\,V^{+-}_{kl}\ .
\end{eqnarray*}

\noindent
{\bf Remark:} The conjecture expressed in terms of \eqr{eq:hirota}, 
 i.e.  that $G(\{T^{+}_k\},\{0\}) =1,$  
is now in a form which resembles the Hirota bilinear
relations, which might be indicating some kind of an integrable
 hierarchy, 
perhaps of Toda-type. 

Because the tau-functions are exponential functions, upon acting on them
by the differential operators, we can factor them out in the expression 
of $G(\{x_k\},\{y_k\})$.  We thus define
        \begin{definition} $P( \lambda\Delta_{+}, \{x_k\},\{y_k\})$ is a
           formal power series in the variables $\lambda\Delta_{+},\,
            \{x_k\}$ and $\{y_k\}$ such that
\[
         G(\{x_k\},\{y_k\}) = P(\lambda\Delta_{+} ,\{x_k\},\{y_k\})\,
                \tau(\lambda\Delta_{+}, \{x_k + y_k\})\,
                \tau(\lambda\Delta_+, \{x_k - y_k\}).
\]
        \end{definition}
Hence, Givental's conjecture for $\QP^1$ can be restated as
        \begin{conjecture}[Givental] The generating function
        $G(\{T^{+}_k\},\{0\})$ is equal to one, or equivalently
\begin{equation}
                P(\lambda\Delta_{+},\{T^{+}_k\}, \{0\}) = {1\over 
                       \tau(\lambda\Delta_{+},\{ T^{+}_k\} )^2}. \label{eq:PT}
\end{equation}
        \end{conjecture}
This conjecture can be verified order by order\footnote{This procedure is
possible because  when $q_0 = q_1=0$, only a finite number of
terms in the free-energies  and their derivatives
are non-vanishing.  In particular,
the genus-0 and genus-1 free energies vanish when $q_0 = q_1 =0$.} 
in $\lambda$.  

Let us check  \eqr{eq:PT} up to order $\lambda^2$, for which we need
to consider up to $\lambda^6$ expansions in the differential
operators acting on the $\tau$-functions.  
Let $h = \lambda \Delta_{+}$.  
The low-genus free energies for a point target space can be easily computed 
using the KdV hierarchy and topological
axioms; they can also be verified using Faber's program 
\cite{Faber}.  The terms relevant to our computation are 

{\small \begin{eqnarray*}
 {{\cal F}^{\mbox{\tiny pt}}_0\over h} + {\cal F}_1^{\mbox{\tiny pt}} + 
 h {\cal F}_2^{\mbox{\tiny pt}} 
 &=& {1 \over h}\left[ {(q_0)^3\over 3!} 
    + {(q_0)^3 q_1\over 3!} 
    + 2! {(q_0)^3(q_1)^2\over 3!\,2!}
    + 3! {(q_0)^3(q_1)^3\over  3!\,3!}
    + {(q_0)^4 q_2  \over 4!} 
    + 3 {(q_0)^4 q_1 q_2\over 4!} 
    +  \right.\\
 && +\, 12 {(q_0)^4 (q_1)^2 q_2\over 4!\,2!} 
    + {(q_0)^5 q_3\over 5!}  + 4{(q_0)^5 q_1 q_3\over 5!} 
    + 6 {(q_0)^5 (q_2)^2  \over 5! 2!}
    + 30{(q_0)^5q_1(q_2)^2\over 5!\, 2!}
    +\\
 && +\left. {(q_0)^6 q_4  \over 6!} 
    +  10{(q_0)^6 q_2 q_3\over 6!} 
    + 90 { (q_0)^6 (q_2)^3 \over 6!\,3!}
    + \cdots \right] + \\   
%           Genus 1
%
&& +  \left[  {1\over 24} q_1 
   + {1\over 24} {(q_1)^2\over 2!}
   + {1\over 12}  {(q_1)^3\over 3!} 
   + {1\over 4} {(q_1)^4\over 4!}
   + {1\over 24} q_0 q_2 
   + {1\over 12} q_0 q_1 q_2 
   + {1\over 4} {q_0(q_1)^2q_2\over 2!} 
   +  \right. \\
&& +\,  {q_0 (q_1)^3 q_2\over 3!} 
  + {1\over 6} {(q_0)^2 (q_2)^2\over 2!\, 2!}
  + {2\over 3} {(q_0)^2q_1 (q_2)^2 \over 2!\,2!}
  + {10\over3} {(q_0)^2 (q_1)^2 (q_2)^2\over 2!\,2!\,2!}
  + {1\over 24} {(q_0)^2 q_3\over 2!} 
  + \\
&&  +\, {1\over 8} {(q_0)^2 q_1 q_3 \over 2!}
  + {1\over2}  {(q _0)^2 (q_1)^2 q_3 \over 2!\,2!} 
  + {7\over 24} {(q_0)^3 q_2 q_3\over 3!}  
  + {35 \over 24} {(q_0)^3 q_1 q_2 q_3\over 3!} 
  + \\
&& +\, 2 {(q_0)^3(q_2)^3\over 3!\,3!}
   + 12 {(q_0)^3 q_1 (q_2)^3 \over3!\,3!}
   +{1\over 24} {(q_0)^3 q_4\over 3!} 
   + {1\over 6} {(q_0)^3 q_1 q_4 \over 3!}
   + 48 {(q_0)^4 (q_2)^4\over 4!\,4!}
   + \\
&& +\left.  {59 \over 12} {(q_0)^4 (q_2)^2 q_3\over 4!\,2!} 
   + {7\over 12} {(q_0)^4 (q_3)^2\over 4!\,2!} 
   + {11\over 24} {(q_0)^4 q_2 q_4\over 4!} 
   + {1\over 24} {(q_0)^4 q_5\over 4!} 
      + \cdots\right] + \\
%
%          Genus 2
%
 &&  +\, h \left[ {7 \over 240}{(q_2)^3\over 3!} 
     +{ 29\over 5760} q_2 q_3 
     +  {1\over 1152} q_4 
     + {7\over 48} {q_1(q_2)^3\over 3!}
     + {7\over 8} {(q_1)^2(q_2)^3\over 2!\,3!} \right. 
     +\\ 
 &&  +\, { 29\over 1440} q_1 q_2q_3 
     + {29 \over 288} {(q_1)^2 q_2 q_3\over 2!} 
     +{1\over 384} q_1 q_4 
     + {1\over 96}{(q_1)^2 q_4\over 2!} 
     + {7\over 12}{q_0(q_2)^4\over 4!} 
     +\\
 &&  +\, { 49\over 12} {q_0q_1 (q_2)^4 \over 4!}
     + {5\over 72} {q_0 (q_2)^2 q_3\over 2!} 
     + {5\over 12} {q_0 q_1(q_2)^2 q_3\over 2!} 
     + {29 \over 2880} {q_0 (q_3)^2\over 2!}
     +\\
 &&  +\,{29\over 576} {q_0 q_1 (q_3)^2\over 2!}
     +{11\over 1440}  q_0 q_2  q_4 
     + {11\over 288} q_0 q_1 q_2 q_4 
     +{1\over 1152} q_0q_5 
     +{1 \over 288} q_0q_1q_5
     + \\
 &&  +\,{ 245\over 12} {(q_0)^2 (q_2)^5 \over 2!\,5!}
     + {11\over 6} {(q_0)^2 (q_2)^3 q_3 \over 2!\,3!}
     + {109\over 576} {(q_0)^2 q_2 (q_3)^2\over 2!\,2!}
     + {17 \over 960 } {(q_0)^2 q_3 q_4\over 2!} + \\
 &&  + \left. {7\over 48 } {(q_0)^2 (q_2)^2 q_4\over 2!\,2!}
     + {1\over 90} {(q_0)^2 q_2 q_5 \over 2!} 
     +{ 1\over  1152} {(q_0)^2 q_6\over 2!} 
     + \cdots\right]\, .
        \end{eqnarray*}}
%%%%%%%%%%%%%%%%%%%%%%%%%%%%%%%%
This expression 
gives the necessary expansion of $\tau(\lambda\Delta_+;\{x_k \pm
y_k\})$ for our consideration, and upon evaluating
$G(\{T^{+}_k\},\{0\})$,
 we find
        \begin{equation}
                P(h, \{T^{+}_k\},\{0\}) = 1  - {17\over 2359296}
        e^{-3t^1/2} h 
        + { 41045 \over 695784701952  } e^{-3t^1}h^2 + {\cal O}( h^3).
        \label{eq:ourP}
        \end{equation}
At this order, the expansion of the right-hand-side of \eqr{eq:PT} is
\[
                \tau( h, \{T^{+}_k\})^{-2}  = 1  - 2\, {\cal
F}_2^{\mbox{\tiny pt}}  
                                \, h + 2 \left[
        ({\cal F}_2^{\mbox{\tiny pt}})^2 \, - {\cal F}_3^{\mbox{\tiny pt}} 
                \right]\, h^2 +{\cal O}( h^3).
\]
At $q_n=T^+_n$, $\forall\, n$, the genus-2 free energy is precisely given by
\[
        {\cal F}_2^{\mbox{\tiny pt}}  = 
                {1\over 1152} T_4  +{29 \over 5760 } T_3 \,T_2  + {7\over 240}
        {T_2^3\over 3!} =  { 17 \over 4718592} e^{-3 t^1/2},
\]
and the genus-3 free energy is 
        \begin{eqnarray*}
        {\cal F}_3^{\mbox{\tiny pt}}  &=& 
                {1\over 82944} T_7  + {77\over 414720} T_2 T_6  + {503\over
        1451520} T_3  T_5  + {17\over 11520}(T_2 )^2 T_5  + {607 \over
        2903040} (T_4 )^2 \\
        &&+
    {1121 \over 241920} T_2 T_3 T_4  + {53\over 6912}  (T_2 )^3 T_4 +
    {583\over 580608}(T_3 )^3 + {205 \over 13824}(T_2 )^2(T_3 )^2 \\
        &&+
    {193\over 6912} (T_2 )^4T_3  + {245\over 20736} (T_2)^6\\
   &=& -\,{656431\over 22265110462464 }\, e^{-3t^1} \, .
        \end{eqnarray*}
Thus, we have
\[
        \tau( h, \{T^{+}_k\})^{-2} = 1-  {17 \over  2359296}\,
e^{-3t^1/2}\, h +
        { 41045 \over 695784701952} \, e^{-3t^1}\, h^2 +{\cal O}(h^3),
\]
which agrees with our computation of $P(\lambda, \{T^{+}_k\},\{0\})$ in \eqr{eq:ourP}.

It would be very interesting if one could actually 
prove Givental's conjecture, but even our particular example remains
elusive and verifying its validity to all orders seems intractable
using our method.

%\vspace{1cm}
\noindent
\section*{Acknowledgments}

\noindent
We would like to thank S. Monni for his collaboration and for
valuable discussions.  J.S.S. also thanks G. Tian and
J. Zhou for useful discussions, and Y.S.S. thanks M. Schulz for helpful
discussions on canonical coordinates.
J.S.S. is  supported in part
by an NSF Graduate Fellowship and the U.S. Department of Energy 
under cooperative research
agreement $\#$DE-FC02-94ER40818.
Y.S.S. is  supported in part by an NSF Graduate Fellowship and the
U.S. Department of Energy under contract \#DE-AC03-76SF00515.

%\newpage
%%%%%%%%%%%%%%%%%%%%


\begin{thebibliography}{999}
        \bibitem{Dubrovin} B. Dubrovin,  Geometry of 2d
        topological field theories, in Integrable systems and quantum
        groups, Lecture Notes in Math. No. 1620 (1996), 120--348.
%
        \bibitem{DZ} B. Dubrovin and Y. Zhang, 
        Bi-Hamiltonian hierarchies in $2D$ topological field theory at
        one-loop approximation.  Comm. Math. Phys. {\bf 198} No. 2
        (1998), 311--361. 
%
       \bibitem{Faber} C. Faber,  Algorithms for computing
        intersection numbers on moduli spaces of curves, with an
        application to the class of the locus of Jacobians, {\it in}
        New trends in algebraic geometry, 93--109, 
        London Math. Soc. Lecture Note Ser., 264,
        Cambridge Univ. Press, Cambridge, 1999.
        % C. Faber, The Maple program accompaniment of
        %{\it Algorithms for computing
        %intersection numbers on moduli spaces of curves, with an
        %application to the class of the locus of Jacobians.}
%
        \bibitem{ellipticGW} A. Givental, Elliptic Gromov-Witten
        invariants and the generalized mirror conjecture,
        Integrable systems and algebraic geometry (Kobe/Kyoto, 1997),
        107--155, 
        World Sci. Publishing, River Edge, NJ, 1998.
        %preprint math.AG/9803053. 
%
        \bibitem{Givental} A. Givental, Semisimple Frobenius
        structures at higher genus, 
        Internat. Math. Res. Notices No. 23 (2001), 1265--1286.
        %preprint math.AG/0008067.
%
        \bibitem{song} J.S. Song,  Descendant Gromov-Witten
        invariants, simple Hurwitz numbers, and the Virasoro
        conjecture for  $\QP^1$. 
        Adv. Theor. Math. Phys. {\bf 3} No. 6 (2000), 1--48.
        %preprint hep-th/9912078.
\end{thebibliography}
\end{document}